\documentclass{PoS}

\usepackage{amsmath}

\newcommand{\beqn}{\begin{eqnarray}}
\newcommand{\eeqn}{\end{eqnarray}}
\newcommand{\eq}[1]{(\ref{#1})}

\newcommand{\Z}{{\mathbb Z}}

\newcommand{\ext}{{\mathrm{ext}}}

\title{Spontaneous electromagnetic superconductivity and superfluidity of QCD$\boldsymbol{\times}$QED vacuum in strong magnetic field}

\ShortTitle{Spontaneous superconductivity of vacuum in strong magnetic field}

\author{\speaker{M. N. Chernodub}${}^{\,}$\thanks{On leave from ITEP, Moscow, Russia. \newline The work of MNC was partially supported by Grant No. ANR-10-JCJC-0408 HYPERMAG.} \\
CNRS, Laboratoire de Math\'ematiques et Physique Th\'eorique, Universit\'e Fran\c{c}ois-Rabelais, F\'ed\'eration Denis Poisson, Parc de Grandmont, 37200 Tours, France\\
Department of Physics and Astronomy, University of Gent, Krijgslaan 281, S9, Gent, Belgium\\
       E-mail: \email{maxim.chernodub@lmpt.univ-tours.fr}}
\author{Jos Van Doorsselaere\\
Department of Physics and Astronomy, University of Gent, Krijgslaan 281, S9, Gent, Belgium\\
       E-mail: \email{jos.vandoorsselaere@ugent.be}}
\author{Henri Verschelde\\
Department of Physics and Astronomy, University of Gent, Krijgslaan 281, S9, Gent, Belgium\\
       E-mail: \email{Henri.Verschelde@ugent.be}}

\abstract{It was recently shown that the vacuum in the background of a strong enough magnetic field may become an electromagnetic superconductor due to interplay between strong and electromagnetic forces. The superconducting ground state of the QCD$\times$QED sector of the vacuum is associated with magnetic--field--assisted emergence of quark--antiquark condensates which carry quantum numbers of charged $\rho$ mesons (i.e., of electrically charged vector particles made of lightest, $u$ and $d$, quarks and antiquarks). Here we demonstrate that this exotic electromagnetic superconductivity of vacuum is also accompanied by even more exotic superfluidity of the neutral $\rho$ mesons. The superfluid component -- despite being electrically neutral -- turns out to be sensitive to an external electric field as the superfluid may ballistically be accelerated by a test background electric field along the magnetic--field axis. In the ground state both superconducting and superfluid components are inhomogeneous periodic functions of the transversal (with respect to the axis of the magnetic field) spatial coordinates. The superconducting part of the ground state resembles an Abrikosov ground state in a type--II superconductor: the superconducting condensate organizes itself in periodic structure which possesses the symmetry of an equilateral triangular lattice. Each elementary lattice cell contains a stringlike topological defect (superconductor vortex) in the charged $\rho$ condensates as well as three superfluid vortices and three superfluid antivortices made of the neutral  $\rho$ condensate. The superposition of the superconductor and superfluid vortex lattices has a complicated ``kaleidoscopic'' pattern.}

\FullConference{Sixth International Conference on Quarks and Nuclear Physics\\
                April 16-20, 2012\\
                Ecole Polytechnique, Palaiseau,  Paris}

\begin{document}

\section{Introduction}

Properties of Quantum Chromodynamics in an extremely strong external magnetic field of hadronic scale have attracted strong interest of the scientific community. For example, hot quark-gluon plasma created in heavy-ion collisions may exhibit the chiral magnetic effect~\cite{Fukushima:2008xe}: chirally--imbalanced matter may generate electric current along the axis of the magnetic field~\cite{Vilenkin:1980fu}. The attractive feature of the chiral magnetic effect is that it may be tested experimentally: the magnetic field of high strength appears naturally in noncentral collisions of heavy ions while the chiral imbalance is generated by topological transitions which are inherent to QCD~\cite{Fukushima:2008xe,Skokov:2009qp}. Moreover, the phase diagram of QCD matter~\cite{Preis:2011sp,Dexheimer:2011pz} is sensitive to the presence of strong magnetic fields which opens a possibility to check these effects experimentally in the astrophysical setup (the magnetic field affects the macroscopic properties of the neutral stars, such as mass, adiabatic index, moment of inertia, and cooling curves~\cite{Dexheimer:2011pz}).

The magnetic field has an impact not only on the properties of the (quark) matter, but it may also affect the quantum vacuum. For example, strong magnetic field enhances the chiral symmetry breaking due to magnetic catalysis phenomenon~\cite{Klimenko:1991he,Gusynin:1994re,Gusynin:1995nb}. Consequently, the background magnetic field changes the finite--temperature phase structure of the QCD vacuum as it shifts the critical temperatures of the confinement--deconfinement and chiral transitions\footnote{One should notice, however, that up to now there is no general consensus in the literature even on a qualitative behavior of these transitions in the background of the strong magnetic field.}~\cite{Gatto:2010pt,Gatto:2010,Mizher:2010zb,D'Elia:2010nq,Fraga:2012fs}.

Another interesting phenomenon is that the sufficiently strong magnetic field may induce a spontaneous {\it electromagnetic} superconductivity of the vacuum~\cite{Chernodub:2010qx,Chernodub:2011mc}. The superconductivity of, basically, empty space, is supported by emergence of specific quark-antiquark condensates with the quantum numbers of $\rho$ mesons (electrically charged vector particles made of lightest, $u$ and $d$, quarks and antiquarks). The spontaneous creation of both charged and neutral  $\rho$--meson condensates should take place if the strength of the magnetic field exceeds the critical value
\beqn
B_c \simeq 10^{16} \, {\mathrm{Tesla}}
\qquad  {\mathrm{or}} \qquad
e B_c \simeq 0.6\,\mbox{GeV}^2\,.
\quad
\label{eq:Bc}
\eeqn 
It is very promising that the magnetic fields of (the twice of) the required strength~\eq{eq:Bc} may be created in the experiments on heavy-ion collisions at the Large Hadron Collider~\cite{Deng:2012}.

Below we discuss the complicated structure of the superconducting ground state in the scope of the effective (``bosonic'') quantum electrodynamics for the charged and neutral $\rho$ mesons~\cite{Djukanovic:2005ag} following Refs.~\cite{Chernodub:2010qx,Chernodub:2011gs}. Similar results can also be obtained in a completely different (``fermionic') approach which uses the Nambu-Jona-Lasinio model~\cite{Chernodub:2011mc}. This somewhat counterintuitive effect (``nothing'' becomes a superconductor) can also be found in holographic approaches~\cite{Callebaut:2011ab,Ammon:2011je} and in numerical simulations of quenched QCD on the lattice~\cite{Braguta:2011hq}. 

Interestingly, the ground state contains also a {\it neutral} superfluid component~\cite{Chernodub:2011tv}. This superfluid turns out to have a rather exotic property as it can be accelerated by an external electric~field.

Another exotic property of this ground state was pointed out recently by Smolyaninov: the superconducting state of the vacuum behaves similarly to a (hyperbolic) metamaterial which may act as diffractionless ``perfect lenses''~\cite{Smolyaninov:2011wc}. Since large magnetic fields may have existed in the early moments of our Universe~\cite{Grasso:2000wj}, a possible metamaterial behavior of the early vacuum may have certain imprints in the large scale structure of the present-day Universe.

\newpage

\section{Electrodynamics of charged and neutral $\rho$ mesons}

The $\rho$ meson is a vector (spin-1) particle made of a light ($u$ or $d$) quark and a light antiquark. There are positively, $\rho^{+} = u \bar d$ and negatively, $\rho^{-} = d \bar u$ charged $\rho$ mesons and a neutral $\rho^{(0)} = (u {\bar u} - d {\bar d})/\sqrt{2}$ meson with approximately the same masses $m_{\rho} \approx 775$\,MeV. A self-consistent quantum electrodynamics for these charged and neutral  vector mesons is described by Djukanovic--Schindler--Gegelia--Scherer (DSGS) Lagrangian~\cite{Djukanovic:2005ag}:
\beqn
{\cal L} & = &-\frac{1}{4} \ F_{\mu\nu}F^{\mu\nu}
- \frac{1}{2} \ \rho^\dagger_{\mu\nu}\rho^{\mu\nu} + m_\rho^2 \ \rho_\mu^\dagger \rho^{\mu}
\label{eq:L:rho}
-\frac{1}{4} \ \rho^{(0)}_{\mu\nu} \rho^{(0) \mu\nu}+\frac{m_\rho^2}{2} \ \rho_\mu^{(0)}
\rho^{(0) \mu} +\frac{e}{2 g_s} \ F^{\mu\nu} \rho^{(0)}_{\mu\nu}\,, \qquad
\eeqn
which generalizes the celebrated vector meson dominance model~\cite{Sakurai:1960ju} by including the electromagnetic (Maxwellian) $U(1)_{\mathrm{e.m.}}$ group:
\beqn
U(1)_{\mathrm{e.m.}}: \quad
\left\{
\begin{array}{lcl}
\rho_\mu(x) & \to & e^{i e \omega(x)} \rho_\mu(x)\,,\\
A_\mu(x) & \to & A_\mu(x) + \partial_\mu \omega(x)\,. \quad
\end{array}
\right.
\label{eq:gauge:invariance}
\eeqn

The Lagrangian~\eq{eq:L:rho} describes the interaction of the electromagnetic field $A_\mu$ with the negatively charged, $\rho_\mu \equiv \rho^- = (\rho^{(1)}_\mu - i \rho^{(2)}_\mu)/\sqrt{2}$, positively charged, $\rho^+_\mu \equiv \rho^\dagger_\mu$, and neutral, $\rho^{(0)}_\mu$, vector mesons, as well as the self-interactions of the mesons. The tensor quantities in Eq.~\eq{eq:L:rho} are various strength tensors,
\beqn
F_{\mu\nu} = \partial_\mu A_\nu-\partial_\nu A_\mu\,, \qquad\qquad\qquad\qquad \ \
\label{eq:F}
{f}^{(0)}_{\mu\nu} & = & \partial_\mu \rho^{(0)}_\nu-\partial_\nu \rho^{(0)}_\mu\,,
\label{eq:f0}\\
\rho^{(0)}_{\mu\nu}
= {f}^{(0)}_{\mu\nu}
- i g_s(\rho^\dagger _\mu \rho_\nu-\rho_\mu\rho^\dagger _\nu)\,, \qquad\qquad
\label{eq:rho0}
\rho_{\mu\nu} & = & D_\mu \rho_\nu - D_\nu \rho_\mu\,,
\label{eq:rho}
\eeqn
and the covariant derivative is $D_\mu = \partial_\mu + i g_s \rho^{(0)}_\mu - i e A_\mu$. The $\rho \pi \pi$ vertex coupling $g_{s}$ is fixed to its phenomenological value $g_s \equiv g_{\rho\pi\pi} \approx 5.88$.

\section{Magnetic--field--induced condensation of charged $\rho$ mesons}

There is a very simple naive argument which supports the condensation of the $\rho$ meson excitations~\cite{Chernodub:2010qx}. Consider a free charged relativistic vector (spin-1) particle. Assume that it has the mass $m$ and the gyromagnetic ratio $g = 2$ (which is strongly supported by various results~\cite{Chernodub:2010qx,Chernodub:2011gs}) . Then in a background of an external magnetic field $\vec B_\ext = (0,0,B_\ext)$ this particle has the following energy spectrum:
\beqn
\varepsilon_{n,s_z}^2(p_z) = p_z^2+(2 n - 2 s_z + 1) eB_\ext + m^2\,.
\label{eq:energy:levels}
\eeqn
where $s_z = -1, 0, +1$ is the spin projection on the field's axis $\hat z \equiv {\hat x}_3$, $n\geqslant 0$ is a nonnegative integer number (which, together with the spin projection $s_z$, labels the Landau levels), and $p_z$ is the particle momentum along the magnetic field's axis $\hat z$. The ground state (given by the quantum numbers $n=0$, $s_z = +1$, $p_z = 0$) has the following (squared) energy: 
$\varepsilon^2(B_\ext) = m^2 - e B_\ext$. The ground energy of the charged $\rho$-meson in the external magnetic field becomes purely imaginary quantity if the magnetic field exceeds the critical value $B_c = m^2/e$, which, in the case of the $\rho$ mesons, corresponds to the critical field~\eq{eq:Bc}. Exactly the same result can be derived from the DSGS model~\eq{eq:L:rho} in the background of the magnetic field\footnote{In the DSGS Lagrangian~\eq{eq:L:rho} the anomalously large gyromagnetic ratio ($g = 2$) of the charged $\rho^\pm$ mesons is described by a nonminimal coupling of the $\rho$ mesons to the electromagnetic field given by the last term in Eq.~\eq{eq:L:rho}.}~\cite{Chernodub:2010qx}. The condensed $\rho$-meson state has the following wavefunction:
\beqn
\rho_1(x_{1},x_{2}) = - i \rho_2(x_{1},x_{2}) = \rho(x_{1},x_{2})\,,
\qquad
\rho_0(x_{1},x_{2}) = \rho_3(x_{1},x_{2}) = 0\,,
\label{eq:wavefunction}
\eeqn
which can be expressed via a single complex scalar function $\rho = \rho(x_{1},x_{2})$. Since the global rotation around the magnetic field axis as well as local electromagnetic gauge transformations~\eq{eq:gauge:invariance} rotate the scalar field $\rho$ by a phase, the $\rho$--meson condensate~\eq{eq:wavefunction} breaks the gauge symmetry\eq{eq:gauge:invariance} spontaneously by ``locking'' the subgroup of the Lorentz rotations with the gauge transformations (``Lorentz gauge locking''~\cite{Chernodub:2010qx}).

Notice that the similar condensation effects exist in other field theories: it is suggested to happen in the standard model of electroweak interactions in a background [with a much stronger than~\eq{eq:Bc}] magnetic field~\cite{Ambjorn:1988tm,Chernodub:2012fi}. Another example is given by a pure Yang-Mills theory, in which the gluons should form a condensate in a {\it chromo}magnetic field background~\cite{Nielsen:1978rm}.

\section{Superconducting ground state: an explicit solution}

The effective bosonic model~\eq{eq:L:rho} in QCD plays the same role as the Ginzburg--Landau model~\cite{ref:Abrikosov} plays in the superconductivity: both bosonic models describe dynamics of the superconducting carriers (the $\rho$--meson excitations in QCD and the Cooper pairs in ordinary superconductivity). Following the standard approach to superconductivity, we solve the $\rho$--meson DSGS model~\eq{eq:L:rho} classically in order to find basic features of the electromagnetic superconductivity of the QCD vacuum. In particular, we utilize the Abrikosov solution in a type--II superconductor~\cite{ref:Abrikosov}, and we choose a general solution of classical equations of motion of the model~\eq{eq:L:rho} in a form of a sum over lowest Landau levels:
\beqn
\rho(z) = \sum_{n \in \Z} C_n h_n\Bigl(\nu, \frac{z}{L_B},\frac{{\bar z}}{L_B}\Bigr)\,, \qquad
\label{eq:phi:z:GL}
h_n(\nu, z, {\bar z}) = \exp\Bigl\{ - \frac{\pi}{2} \bigl(|z|^2 + {\bar z}^2\bigr) - \pi \nu^2 n^2 + 2 \pi \nu n {\bar z}\Bigr\}\,, \qquad
\label{eq:h:z}
\eeqn
where $L_B = \sqrt{2 \pi/(e B)}$ is the magnetic length and $z = x_{1} + i x_{2}$ is a complex coordinate in the transversal (with respect to the direction of the magnetic field axis, $\hat z \equiv {\hat x}_{3}$). Here $\nu$ and $C_{n}$ are certain parameters which are chosen to minimize the corresponding energy functional. A symmetry pattern of the $C_{n}$ coefficients, $C_{n+N} = C_n$, determines the type of the Abrikosov lattice. The solution is generally periodic in the $(x_{1},x_{2})$ plane and it corresponds to a certain lattice symmetry: $N=1$ gives us a square lattice, $N=2$ corresponds to an equilateral triangular lattice etc.

\begin{figure}[!thb]
\begin{center}
\begin{tabular}{cc}
\includegraphics[width=60mm,clip=true]{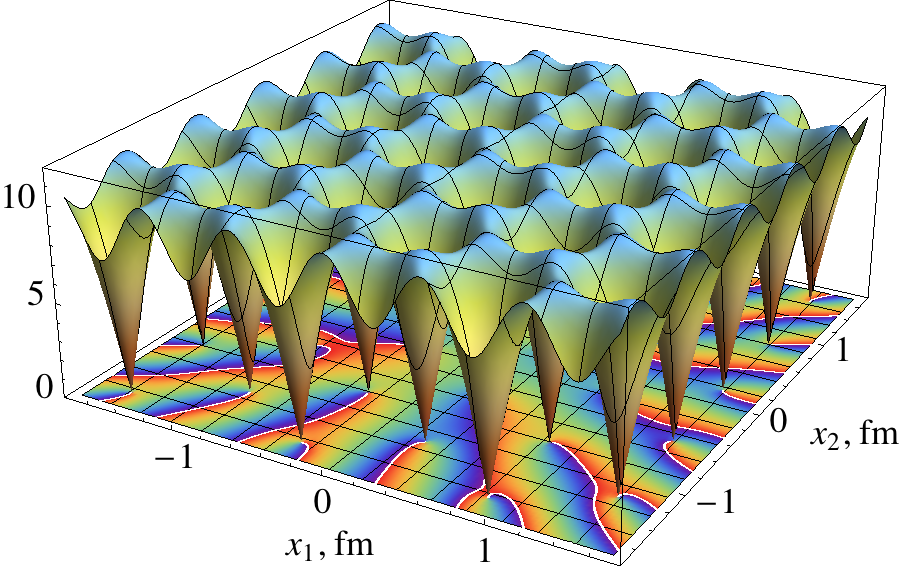} & \\[-37mm]
& \includegraphics[width=80mm,clip=true]{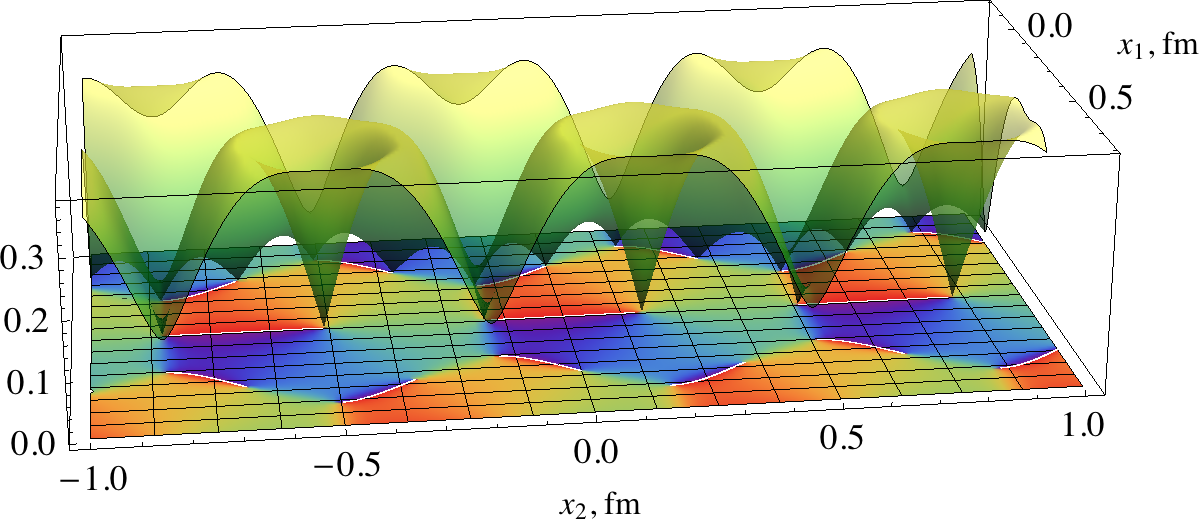} \\
$\rho(x_{1},x_{2})$ & $\rho^{(0)}(x_{1},x_{2})$ 
\end{tabular}
\end{center}
\caption{The charged (left) and neutral (right) $\rho$--meson condensates in the ground state at $B = 1.01 B_{c}$. 
The 3D plots show the behavior of the absolute values (in MeV) of the condensates while the projection on the $(x_{1},x_{2})$ 
plane is a density plot of the phases of the corresponding condensates. The white lines are gauge-dependent singularities 
(Dirac sheets) attached to the superconductor and superfluid vortices, respectively.}
\label{fig:condensates}
\end{figure}

It turns out that in QCD -- contrary to the Ginzburg--Landau model -- the coefficients $C_{n}$ are dependent on the value of the magnetic field, $C_{n} = C_{n}(B)$. In Ref.~\cite{Chernodub:2011gs} it was shown that close to the phase transition, $B \geqslant B_{c}$, the solution is given by the equilateral triangular lattice ($N=2$ with $C_{1} = i C_{2}$ and $\nu = \sqrt[4]{3}/\sqrt{2}$). Each elementary lattice cell contains a stringlike topological defect (superconductor vortex) in the charged $\rho$ condensates as well as three superfluid vortices and three superfluid antivortices made of the neutral  $\rho^{(0)}$ condensate. The superposition of the superconductor and superfluid vortex lattices has a complicated ``kaleidoscopic'' pattern.  The behavior of the charged and neutral condensates in the ground state at $B = 1.01 B_{c}$ are shown in Fig.~\ref{fig:condensates}(left) and \ref{fig:condensates}(right), respectively. Further details can be found in Ref.~\cite{Chernodub:2011gs}.

The electric current density,
\beqn
J_\mu & = & i e \bigl[\rho^{\nu\dagger} \rho_{\nu\mu} - \rho^\nu \rho^\dagger_{\nu\mu} + \partial^\nu (\rho^\dagger_\nu \rho_\mu - \rho^\dagger_\mu \rho_\nu)\bigr] 
- \frac{e}{g_s} \partial^\nu f^{(0)}_{\nu\mu}\,,
\label{eq:Jmu}
\eeqn
satisfies a London--like equation in the longitudinal (parallel to the magnetic field) direction~\cite{Chernodub:2010qx}:
\beqn
\frac{\partial J_3(x)}{\partial x_0}- \frac{\partial J_0(x)}{\partial x_3}  = - \kappa (x_{1},x_{2}) E_3 (x)\,,
\label{eq:London}
\qquad 
\kappa(x_{1},x_{2}) = 4 e^2 \Bigl( \frac{m_0^2}{ - \partial_\perp^2 + m_0^2}  |\rho|^2\Bigr)(x_{1},x_{2})\,,
\label{eq:kappa}
\eeqn
where $\partial_\perp^2 \equiv \partial_{1}^{2} + \partial_{2}^{2}$ is the transversal Laplacian. Relations~\eq{eq:London} constitute the London--type equation which is typical for superconducting systems in which the electric DC current flows without resistance. The quantity $\kappa = \kappa(x_{1},x_{2})$ is a nonlocal function of the superconducting condensate.  In the transversal directions the ground state behaves as an insulator.

The condensation of the charged (superconducting) field $\rho_\mu$ induces condensation of a neutral, superfluid-like field $\rho^{(0)}_\mu$ in the ground state of the vacuum~\cite{Chernodub:2010qx}: 
\beqn
\rho^{(0)}(x_{1},x_{2}) \equiv \rho^{(0)}_{1}(x_{1},x_{2}) + i \rho^{(0)}_{2}(x_{1},x_{2}) = 2 i g_s \cdot \Bigl(\frac{\partial}{- \partial_\perp^2 + m_0^2} |\rho|^2\Bigr)(x_{1},x_{2}) \,,
\label{eq:rho0:explicit}
\eeqn
The longitudinal components of the neutral condensate are zero, $\rho^{(0)}_0 = \rho^{(0)}_3 = 0$.

It is interesting to note that the electrically neutral current of $\rho^{(0)}$ mesons, defined via the relation $J_{\mu}^{(0)} = - \frac{e}{g_s} \partial^\nu f^{(0)}_{\nu\mu}$, satisfies a London--like equation as well:
\beqn
\frac{\partial J^{(0)}_3(x)}{\partial x_0}- \frac{\partial J^{(0)}_0(x)}{\partial x_3}  = - \kappa^{(0)} (x_{1},x_{2}) E_3 (x)\,,
\label{eq:London:rho0}
\qquad 
\kappa^{(0)}(x_{1},x_{2}) = 4 e^2 \Bigl( \frac{\partial_\perp^2}{ - \partial_\perp^2 + m_0^2}  |\rho|^2\Bigr)(x_{1},x_{2})\,.
\quad
\label{eq:kappa:rho0}
\eeqn

Literally, equations \eq{eq:London} and \eq{eq:London:rho0} indicate that if we impose a weak (test) external electric field parallel to the axis of the strong magnetic field, then the charged and neutral currents, respectively, would start to flow frictionlessly along the magnetic field axis. Once the electric field is removed the currents continue to flow because of the absence of the dissipation for the superconducting and superfluid condensates. The sensitivity of the neutral condensate to the electric current is an unexpected feature of this new vacuum ground state.

\end{document}